\documentclass[preprint2]{aastex}
\usepackage{natbib}

\shorttitle{SN1994W}
\shortauthors{Schlegel}

\title{X-ray Detection of SN1994W in NGC 4041?}
    \author{Eric M. Schlegel}
    \affil{Harvard-Smithsonian Center for Astrophysics}
    \affil{Mail Stop 4, 60 Garden Street, Cambridge, MA 02138}
    \email{eschlegel@cfa.harvard.edu}

\begin{abstract}

Optical spectra of SN1994W in NGC 4041 revealed the presence of a
dense (N$_{\rm e}$ $>$10$^8$ cm$^{-3}$ circumstellar shell.  An
observation with the {\it ROSAT} HRI detected a source, with a
luminosity of $\sim$8$\times$10$^{39}$ ergs s$^{-1}$, coincident with
the position of SN1994W to within 1$''$.4.  The positional coincidence
plus the optical evidence for a dense circumstellar shell support the
identification of the X-ray source as SN1994W.

\end{abstract}


\begin{document}

\section{Introduction}

X-ray emission from supernovae arises either from Compton-scattered
${\gamma}$ rays from the radioactive decay of $^{56}$Co or from the
interaction of the circumstellar matter with the supernova's shock
wave.  For the case of circumstellar interaction, X-rays provide a
view of the last years of the progenitor's life, specifically, the
years of mass loss.  Supernovae undergoing circumstellar interaction
are expected to emit X-rays with energies from 1 keV to 100 keV
\citep{CF94}.  The recently-recognized hypernovae class, supernovae
which show evidence of extreme blast wave energies ($\sim$10$^{52}$
ergs s$^{-1}$) \citep{QDW99,FW98,Pac98} and association with GRBs,
also contribute X-ray emission.

Currently, nine ``normal'' supernovae have been detected in the X-ray
band: SN1978K, SN1979C, SN1980K, SN1986J, SN1987A, SN1988Z, SN1993J,
SN1994I, and SN1995N (\citet{Sch95} and references therein for
supernovae earlier than 1994; \citet{Lew95} for SN1995N;
\citet{Imm98a} for SN1979C and \citet{Imm98b} for SN1994I).  Of these,
the early X-ray emission from one (SN1987A) largely arose from
Compton-scattered $\gamma$-rays from the radioactive decay of
$^{56}$Co.  The emission from the other eight is believed to come from
the interaction of the SN shock with a circumstellar medium (the
currently increasing late X-ray emission from SN1987A also comes from
the shock interaction \citep{Has96}).  A tenth supernova, SN1998bw,
identified as a possible hypernova and perhaps associated with
GRB980425, has been detected in X-rays \citep{Pian98}.

This paper describes the detection of X-ray emission from the location
of SN1994W.

\section{Summary of Discovery}

SN1994W in NGC 4041 was discovered by G. Cortini and M. Villi on 1994
July 29.85 \citep{CV94}.  The supernova was located approximately
17$''$.5 N and 7$''$.8 W of the nucleus of NGC 4041 \citep{Pol94}.
\citet{BMB94} obtained a spectrum using the Bologna Astronomical
Observatory 1.5-m that showed a flat continuum with strong H$\alpha$
and H$\beta$ emission lines defining SN1994W as an SN II.  No P Cygni
profiles were observed in the first spectrum. \citet{FB94} reported
that a spectrum obtained with the Lick Observatory 3-m confirmed
SN1994W as a peculiar SN II.  The emission lines showed a narrow
component (FWHM $\sim$1200 km s$^{-1}$) sitting on a broad base (FWHM
$\sim$5000 km s$^{-1}$).  Narrow Fe II emission lines were detected.
Narrow absorption components were visible in the cores of the emission
lines with FWHM $\sim$300 km s$^{-1}$.  Subsequent spectroscopy showed
narrow (FWHM $\sim$1200 km s$^{-1}$) P Cygni profiles.  \citet{CLM94}
described a spectrum obtained about two weeks later that showed little
change in the emission lines.  They suggested that the lack of change
implied the supernova illuminated a dense circumstellar shell.  Radio
and X-ray emission could be expected.

\section{X-ray Observation}

The {\it ROSAT} HRI was used to observe SN1994W on 21-23 October 1997
for an on-source time of 33.7 ksec.  The MJD of the middle of the
observation is 50742.04.  The original processing of the data
contained a 10$''$ boresight error, so the data were re-processed
after a patch was applied to fix the pipeline software.  Sollerman,
Cumming, \& Lundqvist (1998) (hereafter, SCL98) established the date
of outburst as 1994 July 14$^{+2}_{-4}$ so the HRI observation date
corresponds approximately to an age of 1180 days.

The particle background of the data was removed using the software
described by \citet{Snow98}.  The deadtime-corrected exposure totaled
33460.8 sec.  The HRI data were binned to 4$''$ pixels and registered
with an optical image from the Digitized Sky Survey 2
atlas\footnote{Image obtained from http://archive.eso.org/dss/}.  The
binned data were then overlaid on the optical image.  No smoothing was
applied to the X-ray data.  Figure~\ref{overlay} shows the results
with X-ray contours over an optical image.

A point source at the position of SN1994W is visible in the figure.
At the position of the X-ray source, the contours are 2.5, 3.0, and 3.5
counts pixel$^{-1}$.  The 2.5 counts pixel$^{-1}$ contour is
2-3 times the background rate.  The coordinates of the X-ray source
are 12:02:11.0, +62:08:31.7 (J2000) while the coordinates for SN1994W
are 12:02:10.9, +62:08:32.6.  The differences, defined as X-ray minus
optical, are +0$^{\rm s}$.1 in RA and -0$''$.9 in Dec.  To check the
positional accuracy, the coordinates of the nucleus of NGC 4041 were
extracted and compared to published values \citep{Rus90,Pol94}.  The
differences (X-ray minus published) are +0$^s$.1 in RA and +1$''$.1 in
Dec.  The absolute position of SN1994W is accurate to $\sim$1$''$.4, a
value well within the pointing error of the HRI ($\sim$6$''$).

A net total of 31$\pm$7.3 counts was extracted from a 3$''$ radius
circle centered on SN1994W's position.  These counts correspond to a
net source rate of 9.3$\pm$2.2$\times$10$^{-4}$ counts s$^{-1}$.  The
extraction circle contained 50\% of the PSF, so the count rate must be
increased by a factor of 2.  The background that was subtracted was
obtained from an annulus surrounding the galaxy that had an inner
radius of 1$'$ and an outer radius of 2$'$.

The probability of a source falling at the exact location of SN1994W
will be given by the probability of a random background fluctuation at
that location, an estimate based on the log N-log S for the measured
flux, or approximately by the number of sources detected in the galaxy
divided by the galaxy area.  This last number will be the larger of
the three.  For example, from a {\it ROSAT}-measured log N-log S
relation (e.g., \citet{HST91}), one source, of flux $\sim$10$^{-13}$
erg s$^{-1}$ cm$^{-2}$, is expected per square degree (i.e., a
probability of $\sim$2$\times$10$^{-6}$ for the detect cell used
here).  Figure~\ref{overlay} shows at least 4 sources at or above that
flux.  The D$_{25}$ radius of the galaxy is $\sim$2$'$.7
\citep{Tully88}.  If we exclude the $\sim$8$''$ nuclear region, the
resulting annulus has an area of $\sim$7.5$\times$10$^4$ arcsec$^2$ in
which about 8-10 sources are located.  This gives a probability of
$\sim$10$^{-4}$ per arcsec$^{2}$; multiplying by the size of the
detect cell gives a probability of $\sim$3$\times$10$^{-3}$ of a
source falling precisely on the SN1994W position.  We judge this to be
sufficiently small to associate SN1994W with the X-ray source on a
provisional basis.

The Galactic reddening in the direction of NGC 4041 is E$_{\rm B-V}$
$\sim$0.017 \citep{SFD98} which converts to N$_{\rm H}$
$\sim$9$\times$10$^{19}$ cm$^{-2}$ using the conversion of
\citep{PS95}.  However, SCL98 estimate a value for E$_{\rm B-V}$ of
0.17$\pm$0.06 from the interstellar Na I D absorption line using the
calibration of \citet{MZ97}.  That E$_{\rm B-V}$ converts to a column
density of $\sim$9$\times$10$^{20}$ cm$^{-2}$.  We adopt the higher
column because SCL98 detected the Na I line directly toward SN1994W.
The difference between the Galactic and the Na I-determined values for
E$_{\rm B-V}$ is a measure of the absorption local to the SN1994W
region.  Using an assumed thermal bremsstrahlung spectrum with kT
$\sim$5 keV, the count rate converts to an unabsorbed flux in the
0.1-2.4 keV {\it ROSAT} band of $\sim$1.1$\pm$0.3$\times$10$^{-13}$
ergs s$^{-1}$ cm$^{-2}$.  Using a distance of 25.4 Mpc (SCL98), the
L$_{\rm X}$ is 8.5$\pm$2.0$\times$10$^{39}$ ergs s$^{-1}$.

\section{Discussion}

The positional coincidence and the optical spectral behavior argue
that SN1994W has been detected in the X-ray band.  The detection of
narrow absorption lines in the optical spectra requires the existence
of a dense circumstellar shell.  SCL98 estimate a number density for
the shell of $>$10$^8$ cm$^{-3}$ from the presence of optically thick
Fe II lines.  The estimated X-ray luminosity is comparable to that of
SN1986J, SN1988Z, and SN1995N \citep{Sch95} and lies near the upper
end of the X-ray luminosity range for supernovae.

No other X-ray observations have been made of NGC 4041.  The IRAS
observation of NGC 4041 had, at the best spatial resolution, a beam
size of 0$'$.77 which provides essentially no spatial information
\citep{Soi89}.  A VLA image at 4.85 GHz was obtained at 1$'$, again
providing no spatial information \citep{BWE91}.  Certainty that
SN1994W has been detected in the X-ray band will only be possible with
additional observations, particularly with {\it Chandra}.  The X-ray
light curve, whether increasing or decreasing, is of interest.

Table~\ref{tbl-1} lists parameters generated from the competing models
of \citet{CF94} (hereafter, CF94), which applies to normal Type II
supernovae exploding into a red giant wind, and \citet{Ter92}
(hereafter, the `cSNR' model), which describes a supernova expanding
into a very dense circumstellar environment (equations summarized in
\citet{Aret99}).  The parameters for the CF94 model are calculated for
two different values of the power law index $n$ that describes the
expanding gas.  Each model is calculated for two different ages, at a
template value of 30 days and at the age of the {\it ROSAT}
observation.  For the number density, we adopt the lower bound of
3$\times$10$^8$ cm$^{-3}$ (SCL98).

Both models show similar behavior: temperatures and velocities
decrease from day 30 to day 1180 and shell radii and luminosities
increase.  In detail, however, the models differ.  The initial
velocities of the CF94 $n=7$ models are too high; the observed FWZI
velocities near day 30 were $\sim$5000 km s$^{-1}$.  For the $n=20$
model, the initial velocity is correct, but the day 1180 velocity
remains too high and the X-ray luminosity is too low.  The cSNR model
appears to be a better match to the observations.  The luminosities in
the cSNR model are too large by factors of 100-1000, but we have
little or no information for the efficiency of the X-ray production.
An efficiency value of $\sim$0.1-1\% would bring the prediction into
line with the observation.  Optically, SN1994W showed P Cygni profiles
at H$\alpha$ and, for the first 100 days, showed a Type IIP light
curve (SCL98).  The spectrum also resembled the spectra of Type IIn
supernovae such as SN1988Z and SN1987B \citep{Fil97, Sch90, SchK96}.
On the basis of its optical behavior, SN1994W may fall near the middle
of a continuum that has at one extreme the IIn supernovae, with dense
circumstellar shells, and at the other extreme ``normal'' Type II
supernovae with little or no circumstellar medium.  The X-ray
behavior, however, places SN1994W among the most luminous IIn
supernovae (assuming that the distance to NGC 4041 is known
accurately).

We note in passing that the face-on spiral galaxy NGC 4041 is itself
of interest.  At least six point sources have been positively detected
in the HRI image; another six may exist near the nucleus.  The nuclear
emission appears to be extended, although the emission may be the
blended emission of nearby point sources.  The X-ray study of this
galaxy will benefit from an observation with {\it Chandra}.

In summary, an observation of NGC 4041 with the {\it ROSAT} HRI has
revealed the existence of an X-ray source at the position of SN1994W.
The signature of a dense circumstellar shell in the optical spectrum
plus the positional coincidence of the X-ray source with the optical
position support the identification of the X-ray source as SN1994W,
the eleventh supernova discovered to emit X-rays.

\acknowledgements

I thank the referee for comments that improved the presentation.  This
research was supported by NASA Grant NAG5-6923 to the Smithsonian
Astrophysical Observatory.

\begin{deluxetable}{rrrrrrr}
\footnotesize
\tablecaption{Comparison of Model Quantities\tablenotemark{a} \label{tbl-1}}
\tablewidth{0pt}
\tablehead{
\colhead{} &
\colhead{} &
\colhead{} &
\multicolumn{4}{c}{CF94 models} \\
\colhead{} &
\multicolumn{2}{c}{cSNR models} &
\multicolumn{2}{c}{n = 7} &
\multicolumn{2}{c}{n = 20} \\
\colhead{Quantity} & 
\colhead{30 d} & 
\colhead{1180 d} &
\colhead{30 d} &
\colhead{1180 d}  & 
\colhead{30 d} & 
\colhead{1180 d}
}
\startdata
R$_{\rm s}$ (cm)                & 6.9(15) & 1.9(16) & 7.7(15) & 1.4(17) & 1.7(15) & 5.6(16) \\
V$_{\rm s}$ (km s$^{-1}$)       & 7880   & 570 & 31800 & 15300 & 7780 & 6330 \\
T$_{\rm s}$ (K)                 & 8.8(8)  & 4.6(6) & 5.0(8) & 1.2(8) & 1.9(6) & 1.3(6) \\
T$_{\rm s}$ (keV)               & 80  & 0.42 & 45 & 11 & 0.17 & 0.12 \\
L$_{\rm s}$ (ergs s$^{-1}$)     & 1.5(44) & 4.7(41) & 2.9(41) & 3.2(40) & 1.9(39) & 1.0(39) \\
M$_{\rm shell}$ (M$_{\odot}$)   & 0.37    & 8.6 & 2.0(-4) & 3.7(-3) & 4.3(-5) & 1.4(-3) \\
 \enddata
\tablenotetext{a}{Values in parentheses are exponents}

\tablecomments{The adopted quantities applied to both models are as
follows: explosion energy = 1.3$\times$10$^{51}$ ergs s$^{-1}$; number
density of 3$\times$10$^8$ cm$^{-3}$ as a lower limit (SCL98).  For
the CF94 models, the adopted mass loss rate is 10$^{-5}$ M$_{\odot}$
yr$^{-1}$ and a wind velocity of 10 km s$^{-1}$.  The cSNR model
yields a time t$_{\rm sg}$ for the onset of radiative cooling of 18.5
days.}

\end{deluxetable}

\newpage

\figcaption[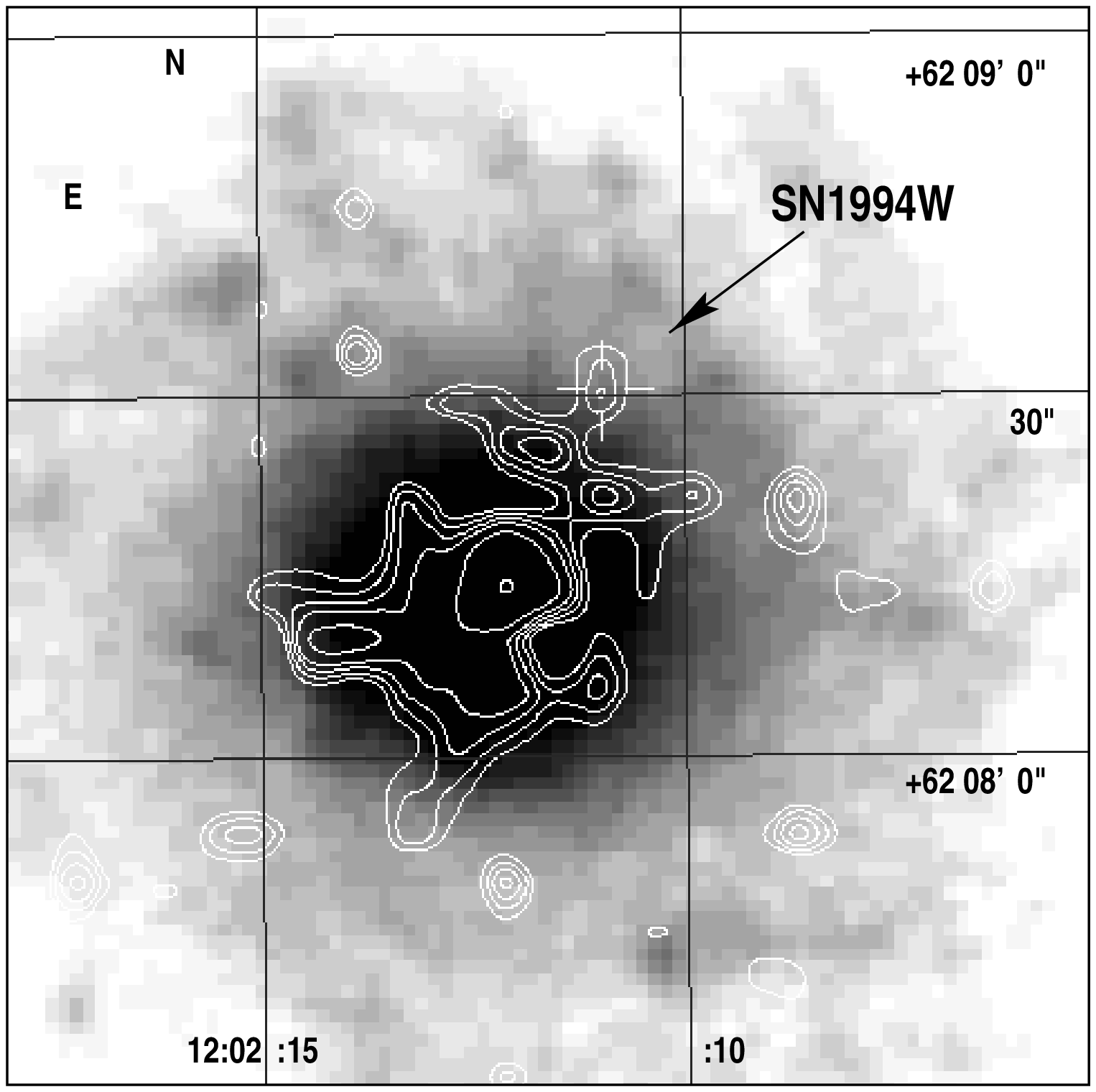]{{\it ROSAT} HRI contours on a DSS2 optical image
of NGC 4041.  The X-ray data were binned to 4$''$ pixels before
registration.  No smoothing was applied to the HRI data before
contouring which explains a few of the ``linear'' features in the
contours. Contours were drawn at 2.5, 3.0, 3.5, 4.0, 5.0, 7.5, 10.0,
12.5, 15.0, and 20.0 counts pixel$^{-1}$.  The contour at 2.5 counts
pixel$^{-1}$ is a factor of 2-3 above the background.  The white `+' marks
the optical position of SN1994W.  \label{overlay}}

\end{document}